\documentclass[reprint,amsmath,amssymb,aps]{revtex4-2}
\usepackage{graphicx}
\usepackage{dcolumn}
\usepackage{bm}
\usepackage{hyperref}
\usepackage[toc,page]{appendix}
\renewcommand\thefigure{\arabic{figure}} 
\usepackage{amsmath,amssymb}

\begin{document}

\preprint{APS/123-QED}


\title{Role of plasticity in the universal scaling of shear thickening dense suspensions}
\author{Sachidananda Barik, Akhil Mohanan}
\author{Sayantan Majumdar}
 \email{smajumdar@rri.res.in}
\affiliation{Soft Condensed Matter Group, Raman Research Institute}

\date{\today}

\begin{abstract}

Increase in viscosity under increasing shear stress, known as shear thickening (ST), is one of the most striking properties of dense particulate suspensions. Under appropriate conditions, they exhibit discontinuous shear thickening (DST), where the viscosity increases dramatically and can also transform into a solid-like state due to shear induced jamming (SJ). The microscopic mechanism giving rise to such interesting phenomena is still a topic of intense research. A phenomenological model proposed by Wyart and Cates shows that the proliferation of stress-activated interparticle frictional contacts can give rise to such striking flow properties. Building on this model, a recent work proposes and verifies a universal scaling relation for ST systems where two different power-law regimes with a well-defined crossover point is obtained. Nonetheless, the difference in the nature of the flow in these two scaling regimes remains unexplored.
Here, using rheology in conjugation with high-speed optical imaging, we study the flow and local deformations in various ST systems. We observe that with increasing applied stress, the smooth flow changes into a spatio-temporally varying flow across the scaling-crossover. We show that such fluctuating flow is associated with intermittent dilatancy, shear band plasticity and fracture induced by system spanning frictional contacts.
\end{abstract}

\maketitle


\section{\label{sec:level1}Introduction}

A classic example of non-Newtonian flow behavior is the increase in viscosity of dense particulate suspensions under increasing shear stress/shear-rate, a phenomenon known as shear thickening (ST) \cite{doi:10.1122/1.550017, doi:10.1063/1.3248476}. Depending on the particle volume fraction ($\phi$), applied stress and nature of interparticle interactions, dense suspensions show a mild or, drastic shear thickening. The mild shear-thickening is known as the continuous shear thickening (CST), while the drastic one is called the discontinuous shear thickening (DST). Interestingly, as $\phi$ approaches the random close packing $\phi_{rcp}$ limit, many of these systems show a stress-induced transformation to a solid-like shear jammed (SJ) state. Such transition has attracted significant recent interest from both fundamental physics as well as, materials design perspectives \cite{peters2016direct, han2016high, Dhar_2020, james2018interparticle, PhysRevLett.124.248005, lee2003ballistic, MAJUMDAR2013191}. As SJ involves a sudden increase in shear modulus as the material transforms to a solid-like state from a liquid-like state, a natural question arises: Is SJ transition a non-equilibrium phase transition? For granular systems, the increase in dynamic correlation reflected in growing length and time scales near the jamming transition also suggest a connection to phase transition and critical phenomenon \cite{PhysRevLett.95.265701}. In the system of soft frictionless spheres, the finite size scaling collapse of elastic moduli and number of contacts further strengthen such idea \cite{Biroli2007PT,PhysRevLett.109.095704}.

For different ST and SJ systems, a very recent work \cite{https://doi.org/10.48550/arxiv.2107.13338} proposes a universal scaling of viscosity over wide range of stress and volume fraction values. This study highlights the importance of two fixed points, namely, isotropic jamming point ($\phi_0$) and frictional jamming point ($\phi_m$) in controlling such scaling behavior. Incorporating the concept of stress induced frictional interaction as proposed in the Wyart-Cates (W-C) model \cite{PhysRevLett.111.218301, PhysRevLett.112.098302, doi:10.1122/1.4890747}, they expressed the viscosity $\eta$ of a shear thickening system as,\\
\begin{align}
    \eta(\phi_0 - \phi)^2 \sim \left(\frac{1}{\phi_0 - \phi_m} - \frac{f(\sigma)}{\phi_0 - \phi}\right)^{-2} 
\end{align}
where $f(\sigma) = e^{-\sigma^*/\sigma}$ denotes the fraction of frictional contacts, $\sigma^*$ is the onset stress for frictional interaction, $\phi_0$ and $\phi_m$ represent the jamming volume fraction without and with the complete frictional interaction, respectively. Such functional form indicates that irrespective of the value of $\phi$, $\eta(\phi_0 - \phi)^2 $ as function of $x_{wc}=\frac{f(\sigma)}{\phi_0 - \phi}$ should diverge at a single point $x_c=\frac{1}{\phi_0 - \phi_m}$ for $\phi < \phi_0$. However, it is observed that $\eta(\phi_0 - \phi)^2 \sim F\left(\frac{f(\sigma)}{\phi_0 - \phi}\right)$ shows a clear variation in diverging point for different $\phi$ values \cite{https://doi.org/10.48550/arxiv.2107.13338}.
Incorporating some careful modifications in Eq. 1, they overcome such discrepancy and derive an universal scaling relation. 

This scaling framework is tested numerically for a ST system of bidisperse particles where a good agreement has been obtained \cite{10.3389/fphy.2022.946221}. Further, the scaling variable can also be modified to incorporate the effect of orthogonal shear perturbations \cite{ramaswamy2022incorporating}. Similar scaling relation has also been used for fractal suspensions showing a direct transition to SJ from a flowing state under increasing stress \cite{D2SM01080A}. Such universal scaling theory seems to be promising approach in combining all the theoretical, numerical and experimental works. The change in power-law slope in the observed scaling curve \cite{https://doi.org/10.48550/arxiv.2107.13338} is attributed to the crossover between two fixed jamming points, $\phi_0$ and $\phi_m$. This suggest that the frictionless and frictional region belongs to two different universality classes and the change in slope reflects a transition from frictionless to frictional regime. Although such picture of stress-induced transition from a frictionless to a frictional regime is interesting, a direct experimental manifestation of such transition on the flow behavior is yet to be explored.  

Here we study the steady state flow behavior of two well characterized dense suspensions using shear rheology and in-situ high resolution boundary imaging. Using particle imaging velocimetry (PIV) technique, we map out spatio-temporal flow and deformation of the system across the crossover of the universal scaling curve. We observe that plasticity plays a major role in changing the slope of the scaling curve.

\begin{figure*}
\includegraphics[scale=.7]{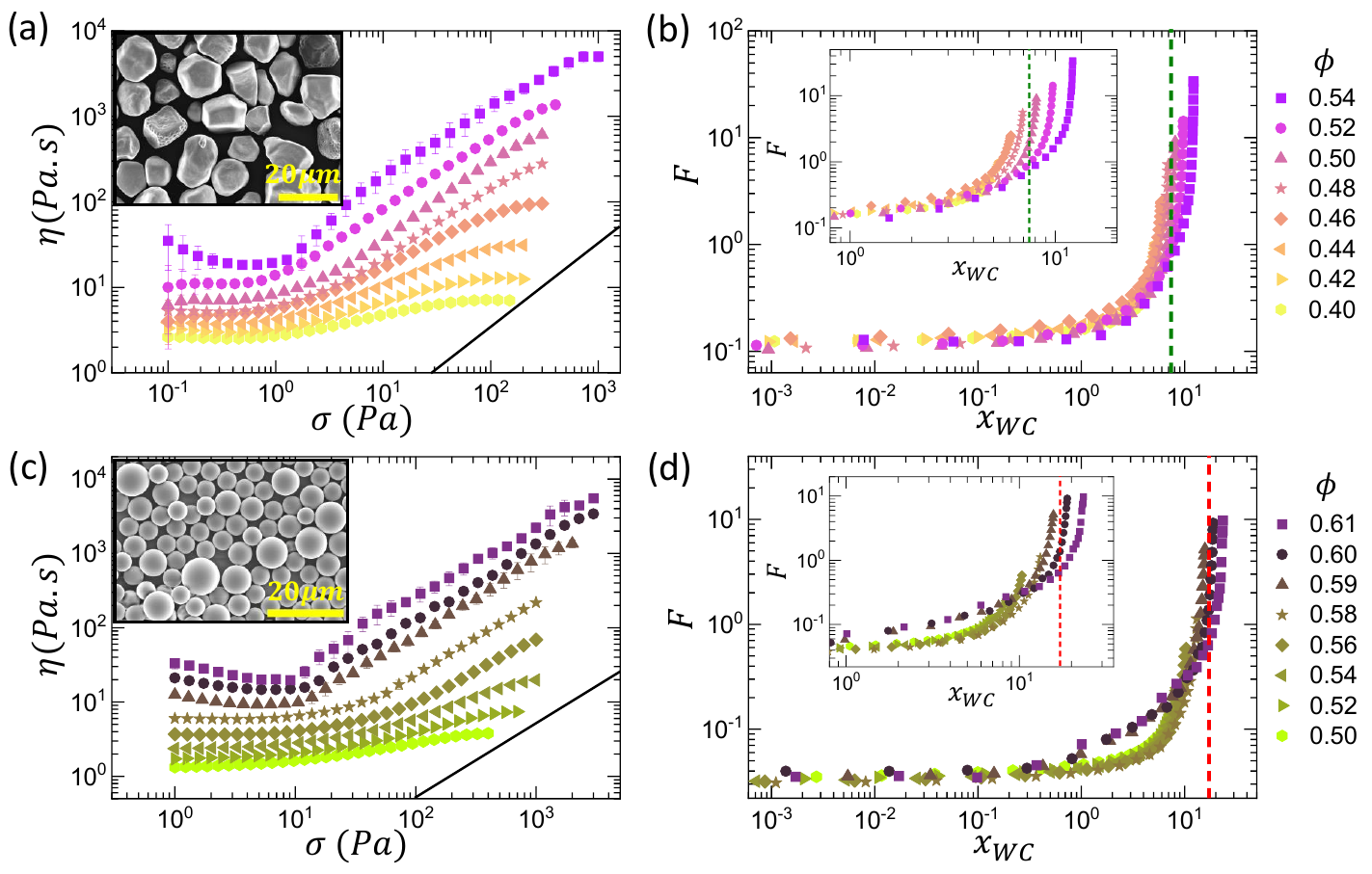}
\caption{\label{fig:wide} (a) and (c) Variation of viscosity $\eta$ with shear stress $\sigma$ shows the ST properties of cornstarch particles dispersed in glycerol and polystyrene particles dispersed in PEG 400 respectively for different volume fraction $\phi$ as shown in the legends. Slope 1 is indicated by the solid black line. The corresponding particle image is represented in the inset. (b) and (d) Variation of WC scaling function $F$ with WC variable $x_{WC}$ for different $\phi$. Inset (magnified plot): $F(x_{WC})$ does not diverge at a single point for different $\phi$. Dashed lines represent the expected point of divergence $x_c = 7.5$ and $17.24$ for cornstarch and polystyrene systems respectively.}
\end{figure*}

 \begin{figure*}
\includegraphics[scale=.53]{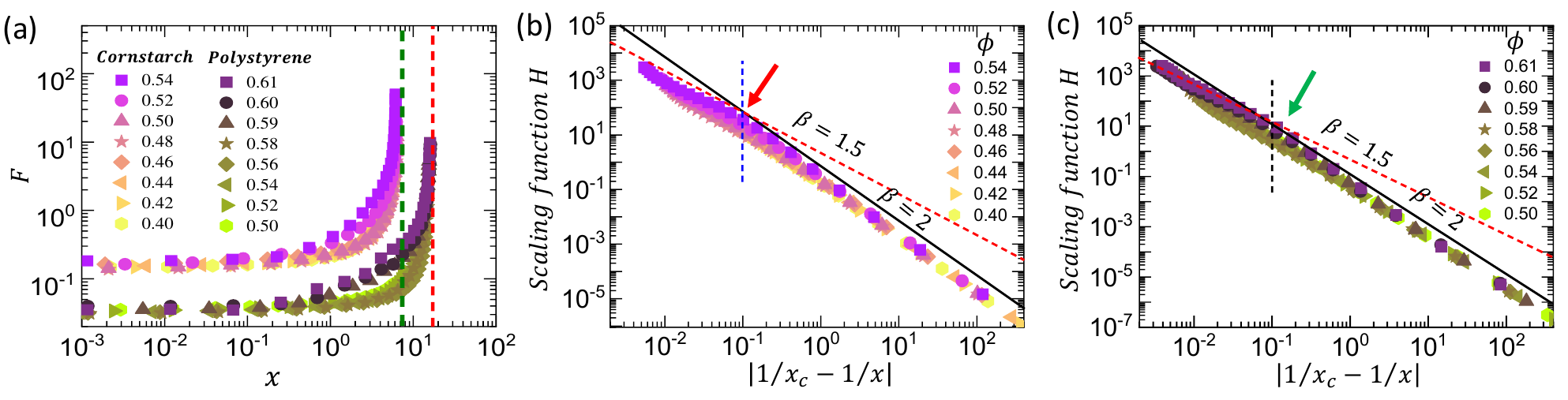}
\caption{\label{fig:wide} (a) Variation of WC scaling function $F$ with modified WC variable $x$ for cornstarch and polystyrene systems at different $\phi$ as shown in legends. Two dashed lines represent the divergence points $x_c$ for the corresponding system. (b) and (c)  Modified scaling function  $ H=\eta(g(\sigma,\phi))^2$ as function of new scaling variable $\left|\frac{1}{x_c}-\frac{1}{x}\right|$ for cornstarch and polystyrene system respectively at different $\phi$. The crossover between the solid and dashed line (marked by the vertical dotted line) represents the change in the magnitude of slope ($\beta$) from $2$ to $1.5$ and the arrow indicates the crossover point.}
\end{figure*}

\begin{figure*} 
\includegraphics[scale=.65]{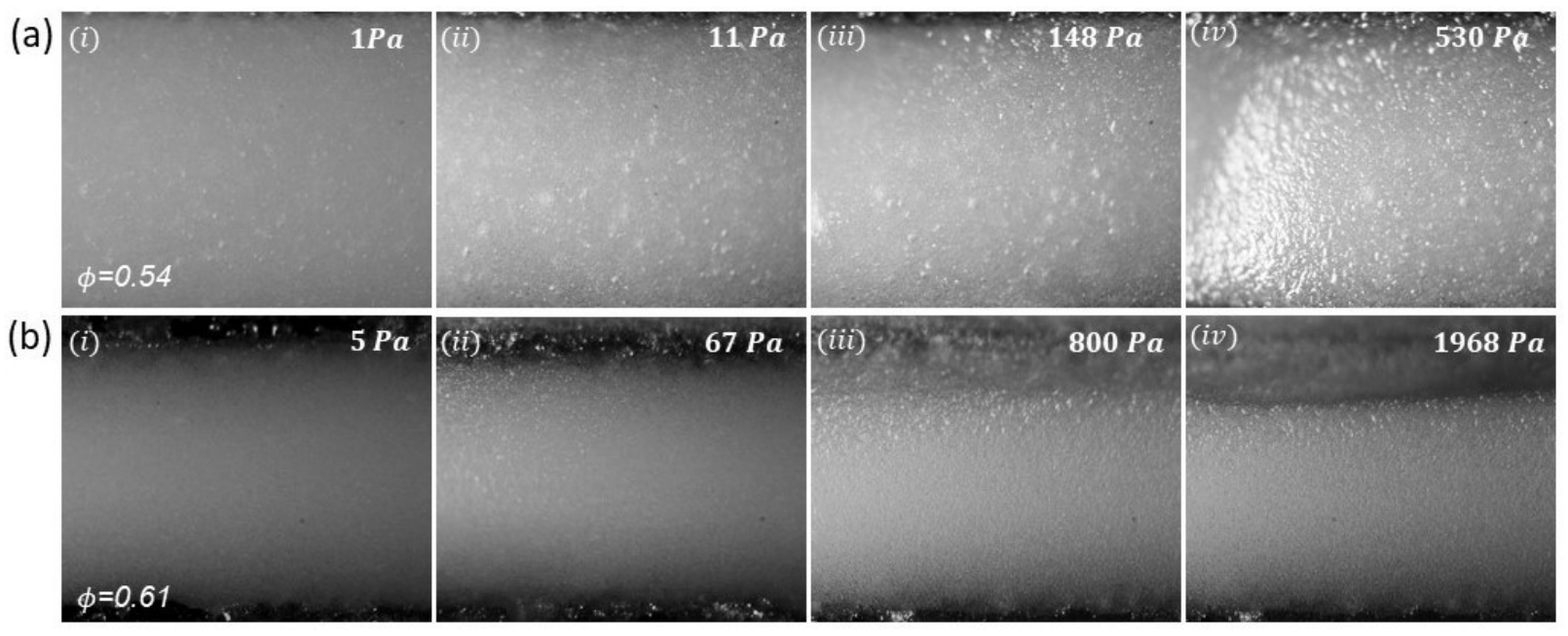}
\caption{\label{fig:wide} Sample boundary images for dense suspension of (a) CS at $\phi = 0.54$  and (b) PS ($d=8 \pm 4$ $ \mu m$) at $\phi = 0.61$ with increasing $\sigma$ during the flow curve. The corresponding $\sigma$ values are mentioned inside the images.}
\end{figure*}

\begin{figure} 
\includegraphics[scale=.52]{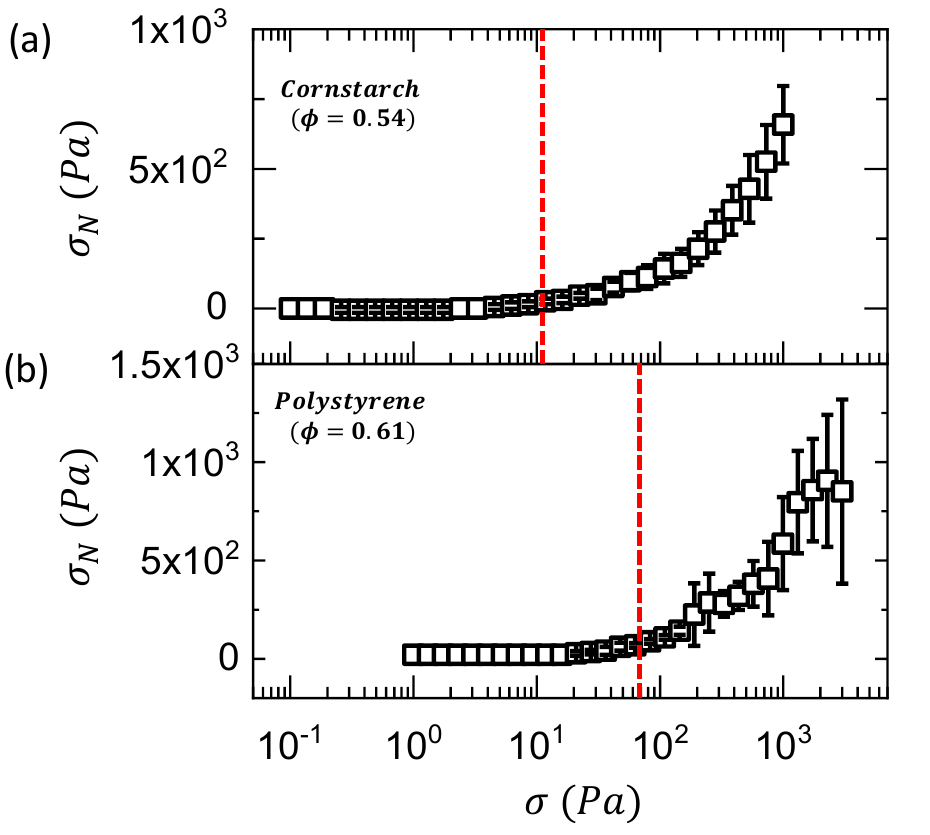}
\caption{\label{fig:epsart}  Variation of normal stress $\sigma_{N}$ with shear stress $\sigma$ during the flow curve for (a) CS system at $\phi = 0.54$ and (b) PS system $\phi = 0.61$. The corresponding crossover point stress $\sigma_{c}$ is marked by the vertical dashed line.}
\end{figure}

\begin{figure*}
\includegraphics[scale=.52]{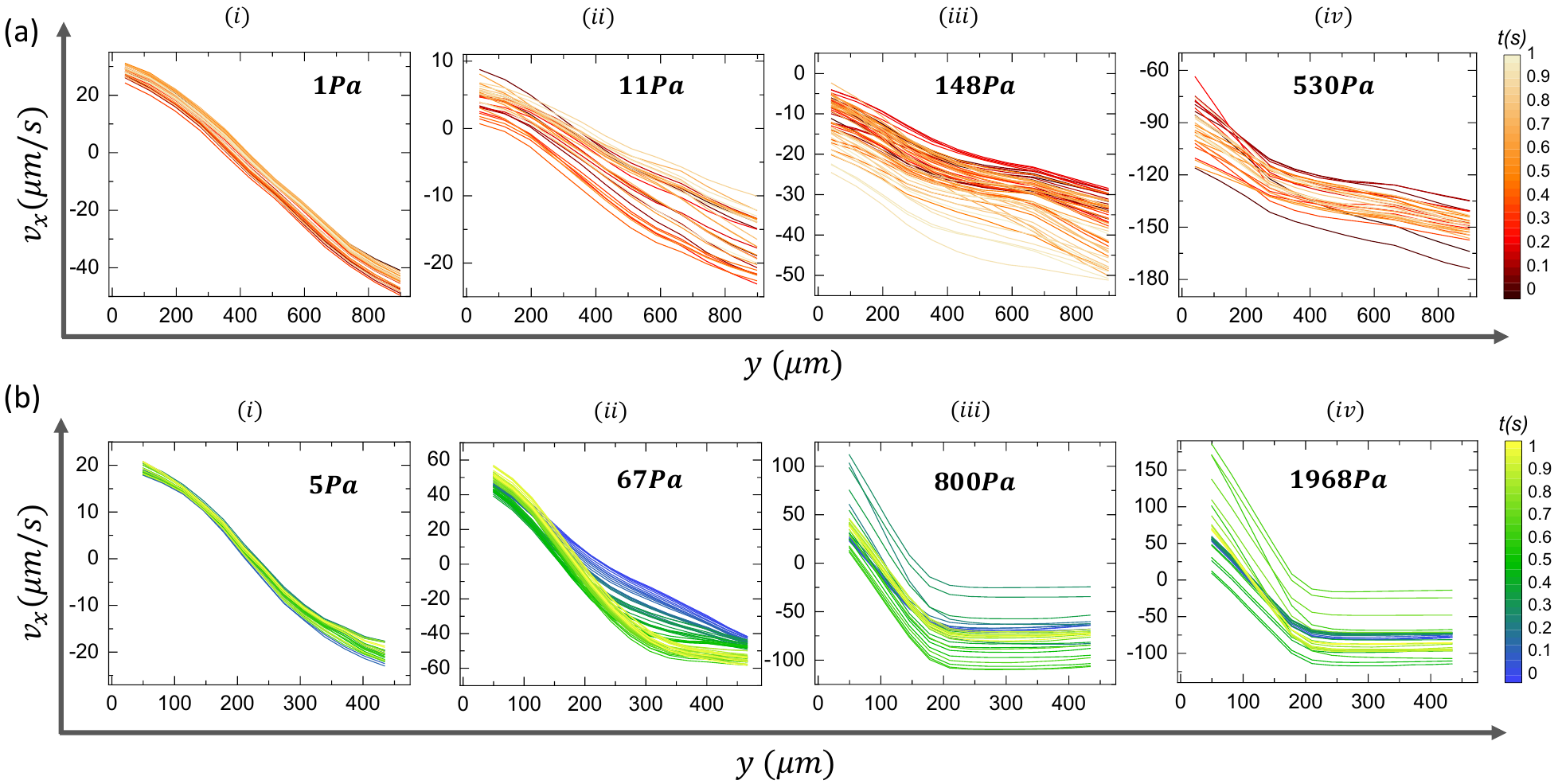}
\caption{\label{fig:wide} Velocity profiles at four different stress values, during the flow curve, for dense suspensions of (a) CS particles at $\phi = 0.54$ and (b) PS particles at $\phi = 0.61$. Each panel represents the velocity profiles for one second, as shown in the color bars, and the corresponding stress values are mentioned inside the panels.  }
\end{figure*}

 \begin{figure*}[hbt!]
\includegraphics[scale=.52]{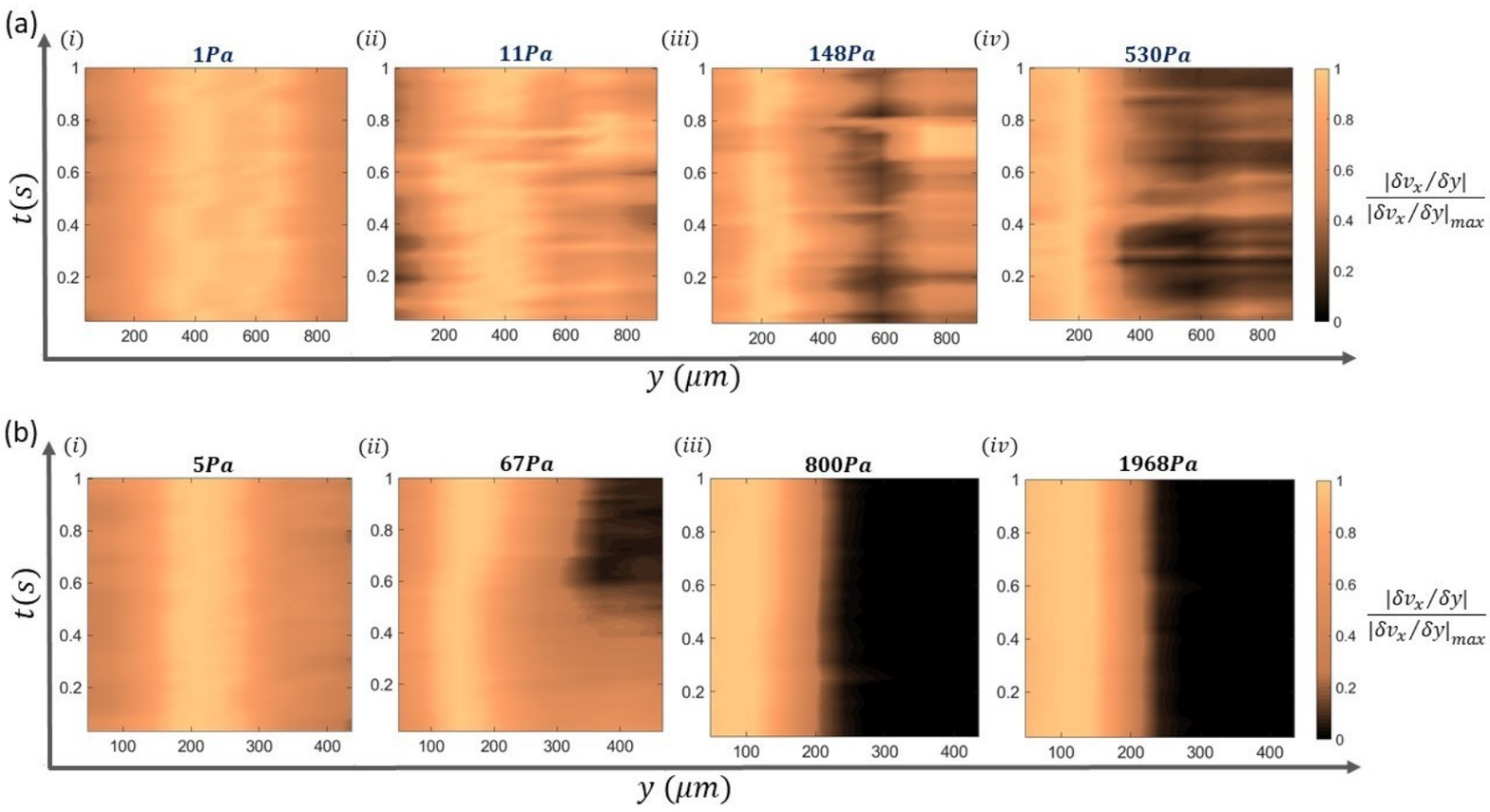}
\caption{\label{fig:wide} Space-time plot (in the y-t plane) of dimensionless velocity gradient ( magnitude of the velocity gradient normalized by the maximum value of gradient magnitude as indicated by the color bar) across the shear gap at four different stress values, as mentioned near each panel, for dense suspensions of (a) CS particles at $\phi = 0.54$ and (b) PS particles at $\phi = 0.61$.}
\end{figure*}

\section{\label{sec:level1}Materials and Methods}
The experiments are carried out for two standard shear thickening (ST) dense suspensions,(i) Cornstarch (CS) particles (Sigma-Aldrich) with mean diameter $d=15\pm 5$ $ \mu m$ dispersed in glycerol and (ii) Polystyrene (PS) particles dispersed in polyethylene glycol(PEG) 400. We synthesize PS particles of two different size distributions ($d=8 \pm 4$ $\mu m$ and $d=2.65 \pm 0.13 $ $\mu m$) using the dispersion polymerization technique \cite{Dhar_2020, PhysRevLett.128.258002}. The samples of different volume fractions ($\phi$) are prepared by gradually adding the required amount of dry particles into the solvent and thoroughly mixed at each step using a spatula. The CS suspensions are then immediately used for rheological measurement. In the case of of PS particles, the suspensions are desiccated overnight to remove air bubbles trapped inside. To make sure an uniform dispersion, the samples are ultrasonicated for 5 minutes (at room temperature) just before loading for rheological measurements. The volume fraction range used for CS system is $0.40 \leq \phi \leq 0.54 $ and that for PS system is $0.50 \leq \phi \leq 0.61$.

All the rheological measurements are performed on a stress control Rheometer (MCR-702, Anton Paar, Austria) in twin drive 50-50 counter movement mode. For the CS particle suspension the measurements are carried out using parallel plate geometry of plate diameter $25$ $ mm$ with $1 $ $mm$ gap between the plates and for PS particle suspensions cone and plate geometry with the cone angle of $2$\textdegree and cone/plate of $25$ $ mm$ diameter is used. In all cases, the surface of the geometries is sandblasted to minimize the wall slippage. In order to remove the loading history, if any, we conduct an oscillatory amplitude sweep measurement at a frequency of 1 $Hz$ just after loading the sample. For this, we first increase the strain amplitude ($\gamma$) logarithmically from a lower to a maximum value and then the amplitude is again gradually decreased to the initial value. For steady state flow curve measurements, we increase the applied shear stress ($\sigma$) from a smaller to maximum value logarithmically with waiting time per data point varying between $20$ $ s$ to $1$ $s$ between minimum and maximum stress, respectively, and then perform a reverse run.


Using a high-speed in-situ imaging set up we obtain the sample boundary images during the flow curve measurement. Since the samples are optically opaque, the in-situ imaging of the boundary is possible only under reflection mode \cite{chattopadhyay2022effect}. The sample boundary is illuminated using a LED light source (Dolan-Jenner Industries) and image the flow gradient plane using a high-speed monochrome CMOS camera (Phantom Miro C210) with $5X$ and $10X$ long working distance objective (Mitutoyo). All the images are captured at a frame rate of $100 $ $Hz$ with resolution $1280\times1024$ and $1280\times720$ pixels for parallel plate and cone-plate geometries, respectively.

From the sample boundary images, we map out the flow profiles using particle imaging velocimetry (PIV) technique. The PIV analysis is performed using custom-written PIV codes developed in MATLAB software.

\section{\label{sec:level1} RESULTS AND DISCUSSION}

The flow curves are obtained for dense suspensions of cornstarch and polystyrene at different volume fractions ($\phi$) as shown in Fig.1(a) and 1(c) respectively. We observe an increase in viscosity ($\eta$) with increasing shear stress ($\sigma$) beyond an onset stress indicating shear thickening. The degree of shear thickening enhances with increasing $\phi$. The black solid lines in Fig.1(a) and 1(c) having a power-law slope of 1 indicate that CS and PS systems show DST for $\phi \geq 0.48$ and $\phi \geq 0.58$, respectively. The respective onset stress values for CS and PS systems are close to 1 Pa and 10 Pa. From the flow curves, using Krieger-Dougherty (KD) equation and WC model we determine $\phi_0$, $\phi_m$, and $\sigma^*$ for both of these systems to estimate the WC scaling variable and scaling function mentioned in \cite{https://doi.org/10.48550/arxiv.2107.13338} (details are given in Appendix C1). Here, we note that the $\sigma^*$ value obtained from W-C Model is higher than the onset stress for shear thickening \cite{Dhar_2020}. Fig.1(b) and 1(d) show the variation of WC scaling function $F=\eta(\phi_0 - \phi)^2$ with WC variable $x_{WC} = \frac{f(\sigma)}{\phi_0 - \phi}$. It shows that for different $\phi$,  $\eta(\phi_0 - \phi)^2\sim F\left(\frac{f(\sigma)}{\phi_0 - \phi}\right)$ diverges at different points (inset of Fig.1(b) and Fig.1(d)), whereas, from equation 1, it is expected to diverge at $x_c=\frac{1}{\phi_0 - \phi_m}$, as marked by the dashed lines.
In order to avoid this discrepancy, following \cite{https://doi.org/10.48550/arxiv.2107.13338}, Eq. 1 can be modified as, \\
 \begin{align}
     \eta(\phi_0 - \phi)^2\sim\left(\frac{1}{\phi_0 - \phi_m} - \frac{g(\sigma,\phi)}{\phi_0 - \phi}\right)^{-2}
 \end{align}
 where $g(\sigma,\phi)= C(\phi)f(\sigma)$ and $C(\phi)$ is a volume fraction dependent parameter called the anisotropy factor. Now plotting $\eta(\phi_0 - \phi)^2 $ as function of $x=\frac{g(\sigma,\phi)}{\phi_0 - \phi}$ gives an excellent collapse at the diverging point $x_c=\frac{1}{\phi_0 - \phi_m}$ [Fig.2(a)]. We would like to point out that here we incorporate only the essential modification required to collapse the diverging points \cite{https://doi.org/10.48550/arxiv.2107.13338}.
 
 Eq.2 suggests that, $\eta(\phi_0 - \phi)^2\sim F\left(\frac{g(\sigma,\phi)}{\phi_0 - \phi}\right)$ should diverge with exponent -2. For better visualization of the exponent, Eq.2 can be recast as (see Appendix A), \\ 
 \begin{align}
     \eta(g(\sigma,\phi))^2 \sim (\phi_0 - \phi_m)^2\left(\frac{1}{x}-\frac{1}{x_c}\right)^{-2}\\
     \eta(g(\sigma,\phi))^2 \sim H\left(\bigg|\frac{1}{x_c}-\frac{1}{x}\bigg|\right)
 \end{align}
 Plotting the scaling function $ H=\eta(g(\sigma,\phi))^2$ as a function of scaling variable $\left|\frac{1}{x_c}-\frac{1}{x}\right|$, we can see that the viscosity $\eta$ over a wide range of $\sigma$ and $\phi$ collapse into a single curve, called universal scaling curve as shown in Fig.2(b) and 2(c) for CS and PS systems, respectively. Interestingly, we observe that magnitude of the slope of the scaling curve ($\beta$) that represents the exponent  of viscosity divergence, does not remain the same: $\beta$ decreases from 2 below a certain value of scaling variable and approaches 1.5. Here we note that the smaller value of the scaling variable $\left|\frac{1}{x_c}-\frac{1}{x}\right|$ corresponds to the larger value of $\sigma$ and vice versa. This suggests the change in the slope of the universal scaling curve happens beyond a certain value of the $\sigma$ called the crossover point stress ($\sigma_{c}$). The crossover points, where the slope change starts, are marked by the red and green arrows and the $\sigma_{c}$ for the CS system ($\phi=0.54$) and PS system ($\phi=0.61$) are $11$ Pa and $67$ Pa, respectively. The change in magnitude of the power-law slope from 2 to 1.5 is particularly evident for higher $\phi$ values (Fig.2(b) and 2(c)), for which the flow curves can be obtained for larger stress values before the sample yields, as shown in Fig.1(a) and 1(c).
 
 In order to understand the physical origin of the change in slope in the universal scaling curve, we use in-situ sample boundary imaging during the rheological measurements \cite{article, Baksi, SebaniJCP, PhysRevLett.128.258002}. Fig.3 represents the boundary images of the sample at different applied $\sigma$ (marked inside each panel) for CS (Fig.3(a)) and PS (Fig.3(b)) system for $\phi = 0.54$ and $\phi = 0.61$, respectively. During the flow curve measurement, we observe no change in surface intensity till $\sigma < \sigma_{c}$ (panel (i) of Fig.3(a) and 3(b)). However, for $\sigma > \sigma_{c}$ significant increase in the surface intensity is observed (panel (ii) and (iii) of Fig.3(a) and 3(b)). Such change in intensity arises due to the shear induced protrusion of the particle at the suspension-air interface due to a mechanism called frustrated dilation \cite{Dilation1, Dilation2}. Shear-induced dilation in dense granular suspensions is related to the shear-induced proliferation of interparticle frictional contacts spanning the system \cite{Dilation1, Dilation2, Dilation3, PhysRevLett.128.258002, smith2010dilatancy, metzner1958flow, doi:10.1021/la000050h}. The dilation starts at $\sigma \approx \sigma_{c}$ (panel (ii) of Fig.3(a) and 3(b), Appendix movie 2 and 3). Similar dilation effect has been reported earlier for granular systems  in the shear thickening regime \cite{Dilation1, Dilation2}. We find from the imaging experiments that for $\sigma \approx \sigma_{c}$ the dilation is intermittent in nature [Appendix movie 1]. However, as $\sigma$ goes beyond $\sigma_c$ the dilation becomes stable and system-spanning. On further increase in $\sigma$, we observe the development of failures in the system in the form of boundary fracture or shear-band plasticity (panel (iv) of Fig.3(a) and 3(b)).

To further characterize the flow regimes, we study the variation of normal stress in the system under shear. For smaller applied stress values below $\sigma_{c}$ the normal stress ($\sigma_{N}$) remains negligible, however, after the crossover point ($\sigma \geq \sigma_{c}$) a significant positive $\sigma_{N}$ starts to develop that increases with increasing $\sigma$ (Fig.4(a) and 4(b)). Such behavior of normal stress, together with the appearance of dilation confirms the  frictional nature of the flow above $\sigma_c$ \cite{PhysRevLett.116.188301, Dilation1, Dilation2, PhysRevLett.128.258002}. Here we observe that the onset of dilation and positive normal stress coincides with $\sigma_c$. On the other hand, dense suspensions of smaller polystyrene particles (closer to colloidal regime) require very high stresses $\sigma >> \sigma_c$ for the observation of dilation and positive $\sigma_{N}$ (Appendix C2). This is due to the high values of confining stresses (proportional to the inverse of particle size) \cite{doi:10.1122/1.4709423}.

Next, we quantify the flow dynamics in the different scaling regimes using Particle Image Velocimetry (PIV) technique. For a given applied stress ($\sigma$) we measure the average value of the velocity components ($v_x$) parallel to the plate motion. The variation of $v_x$ across the gap ($y$) between the shearing plates gives the velocity profiles in the flow-gradient plane. Here, as per our convention for the twin drive 50-50 counter movement mode, the $v_x$ close to the top plate ($y$ = 0) is positive, and the bottom plate is negative  (Appendix C3). Fig.5 represents the evolution of the velocity profile with increasing $\sigma$ for both corn starch (CS) and polystyrene (PS) suspensions (Fig.5(a) and 5.(b)) for $\phi = 0.54$ and $\phi = 0.61$, respectively. For $\sigma < \sigma_{c}$, as shown in panel (i) of Fig.5(a) and 5(b), the velocity profiles remain stationary in time and the nature remains close to a linear profile. The position of the zero velocity plane ($v_x = 0$) appears near the middle of the gap between the two plates. Interestingly, when $\sigma$ approaches $\sigma_{c}$ the velocity profiles lose the stationary behavior showing random shifting in time (Panel (ii) in Fig.5(a) and 5(b)). Here, the zero velocity plane starts shifting towards one of the plates. Similar to the dilation event at $\sigma \approx \sigma_{c}$, the velocity profiles are also spatio-temporally fluctuating in this stress regime. However, they become more stable with increasing $\sigma$. At $\sigma >> \sigma_{c}$ (Panel (iii) and (iv) in Fig.5(a) and 5(b)) we observe  shear band plasticity and fracture in the system (Appendix Movies 2 and 3). We find that the nature of shear banding may differ from system to system. For the CS system (Panel (iii) and (iv) in Fig.5(a)), the entire sample moves in the direction opposite to the direction of motion of the top plate, whereas, the PS system shows two band structures moving opposite to each other (Panel (iii) and (iv) in Fig.5(b)). The density matching of the system (CS is not density matched with the solvent but, PS is) might play a role in creating such differences.

We now quantify the spatio-temporal fluctuations in the velocity profile with increasing applied stress, as mentioned above. Fig.6 shows the space-time plot (STP) of the magnitude of the normalized velocity gradient across the shear gap. The maximum magnitude of the gradient corresponding to a particular applied stress ($\sigma$) is considered to be the normalization factor. Thus, in our case the normalized gradient always remains between 0 and 1. We find that for the lower value of $\sigma$ ($< \sigma_{c}$), the spatio-temporal fluctuations of velocity gradient are negligible, as evident from the uniform colour in panel(i) of Fig.6(a) and 6(b). For $\sigma \approx \sigma_{c}$ when $\sigma$ approaches 11 Pa and 67 Pa in CS and PS  systems, respectively, the spatio-temporal fluctuations get significantly enhanced as reflected by the appearance of patterns in the STP (panel(ii) of Fig.6(a) and 6(b)). For sufficiently high $\sigma$ ($>> \sigma_{c}$), the spatio-temporal fluctuation increases further leading to the shear band and plasticity and fracture in the system (panel(iii) and (iv) of Fig.6(a) and 6(b)). These observations point out that as the frictional contacts proliferate in the system with increasing $\sigma$, the flow becomes more and more heterogeneous leading to material failure. As mentioned earlier, the details of the plasticity and failure depends on the nature of the sample. We find that for CS system, the enhancement of fluctuations continues above $\sigma_c$. However, for PS system, just above $\sigma_c$ we find a clear fracture when the sample splits into two different parts moving in two opposite directions with the shearing plates. Each part has negligible gradient as shown in panel (iii) and (iv) of Fig.6(b). Such differences are more clearly visible in the Appendix movies 2 and 3.  

\section{\label{sec:level1} Conclusions}

In summary, we study the spatio-temporal flow behavior of two well-studied shear-thickening granular suspensions, in relation to the recently reported universal scaling behavior in shear-thickening systems. Similar to the earlier study, we also observe the power-law scaling in these systems with a cross-over point marking a stress-induced transition from frictionless to frictional jamming regime.
Remarkably, our in-situ boundary imaging reveals clear signatures of frictional flow, such as, dilation, positive normal stress, above the critical applied stress corresponding to the scaling cross-over. Thus, our study experimentally verifies the transition from lubrication to frictional regime as predicted by the universal scaling theory for shear-thickening systems \cite{https://doi.org/10.48550/arxiv.2107.13338, 10.3389/fphy.2022.946221}. However, we find that such one to one correlation exists only for granular suspensions. Close to colloidal regime, significant dilation takes place for stress values much beyond the onset stress for shear-thickening due to higher confining stresses in these systems. Thus, from boundary imaging, visible signature of plasticity and scaling cross-over point do not match in this case [Appendix C2]. Deciphering the role of plasticity in controlling the scaling cross-over point for colloidal systems remains an interesting future direction to explore. Our observation highlights the role of plasticity in reducing the slope of the scaling curve from -2 towards -1.5. Such change in slope may be linked to the the plasticity induced softening of viscosity divergence in shear jammed dense suspensions \cite{Dhar_2020}, where the viscosity-divergence exponent close to jamming is given by a Krieger-Dougherty type relation. However, the generality of such connection needs be tested for a wider range of shear-thickening systems.

\section{\label{sec:level1} Acknowledgments}

S.M. thanks SERB (under DST, Govt. of India) for a Ramanujan Fellowship. We acknowledge Ivo Peters for developing the MATLAB codes used for PIV analysis, K.M. Yatheendran for helping with the SEM imaging. S.M. thanks A.K. Sood for helpful discussion.

\providecommand{\noopsort}[1]{}\providecommand{\singleletter}[1]{#1}%
%


\appendix

\section{}

Using Wyart-Cates (W-C) model the viscosity $\eta$ of a shear thickening system can be expressed as\\
\begin{equation} \label{eq1}
    \eta(\phi_0 - \phi)^2 \sim \left(\frac{1}{\phi_0 - \phi_m} - \frac{f(\sigma)}{\phi_0 - \phi}\right)^{-2} 
\end{equation}
\newline
where $f(\sigma) = e^{-\sigma^*/\sigma}$ denotes the fraction of frictional contacts, $\sigma^*$ is the onset stress for frictional interaction, $\phi_0$ is the isotropic jamming point and $\phi_m$ is the frictional jamming point representing the jamming volume fraction without and with the complete frictional interaction, respectively.

After incorporating careful modifications, such as, multiplying $f(\sigma)$ with $C=\left(\frac{\frac{1}{\phi_0 - \phi_m }}{\frac{f(\sigma)}{\phi_0 -\phi}}\right)$, the equation (A1) becomes:

\begin{equation} \label{}
    \eta \sim (\phi_0 - \phi)^{-2}\left(\frac{1}{\phi_0 - \phi_m} - \frac{Cf(\sigma)}{\phi_0 - \phi}\right)^{-2}
\end{equation}

\begin{equation*} \label{}
    \eta \sim \left(\frac{\phi_0 - \phi}{\phi_0 - \phi_m} - {Cf(\sigma)}\right)^{-2}
\end{equation*}

Considering $ g(\sigma,\phi)=Cf(\sigma)$
\begin{equation*} \label{}
\eta (g(\sigma,\phi))^2 \sim \left(\frac{1}{Cf(\sigma)}\frac{\phi_0 - \phi}{\phi_0 - \phi_m} - 1\right)^{-2}
\end{equation*}
Taking common $\frac{1}{\phi_0 -\phi_m}$
\begin{equation} \label{}
\eta (g(\sigma,\phi))^2 \sim (\phi_0 -\phi_m)^2 \left(\frac{\phi_0 - \phi}{Cf(\sigma)} - (\phi_0 - \phi_m)\right)^{-2}
\end{equation}
\begin{equation*}
\sim (\phi_0 -\phi_m)^2\left(\frac{1}{x} - \frac{1}{x_c}\right)^{-2}
\end{equation*}
 \begin{equation}
     \eta(g(\sigma,\phi))^2 \sim H\left(\bigg|\frac{1}{x_c}-\frac{1}{x}\bigg|\right)
 \end{equation}
\vspace{1 cm}
\section{Movie description}
Movie 1 explains the intermittent nature of the dilation around the crossover point stress ($\sigma_c \approx 60$ $Pa$) for polystyrene ($d=8 \pm 4$ $\mu m$) dense suspension at $\phi = 0.58$. The dilation phenomena are reflected in the increase in brightness of the sample boundary image. As the dilation is intermittent in nature around $\sigma_c$, this brightness disappears under shear and again reforms.

Movies 2 and 3 explain the flow and deformation behavior of cornstarch dense particulate suspension at $\phi = 0.54$  and polystyrene ($d=8 \pm 4$ $\mu m$) dense particulate suspension at $\phi = 0.61$ respectively. In movies 2 and 3, we have shown the sample boundary images and corresponding velocity vector images simultaneously for three different stress values at three different regimes of the scaling curve, (i) before the crossover point (ii) during the crossover point (iii) after the crossover point. For the lower stress, before the crossover point, the flow behavior of the sample is almost Newtonian as observed from the velocity vector images, and no change in sample surface intensity is observed. But during the crossover point stress ($\sigma_c$) the flow profile starts distorting in addition to the visible change in sample surface intensity due to dilation. At higher stress, beyond the crossover point, plasticity starts dominating. 

All the images are captured using a CMOS camera (Phantom Miro C210) with a $5X$ long working distance objective (Mitutoyo). The images are captured at a frame rate of 100Hz with resolution $1280 \times720$ $pixels$ for movies 1 and 3, and $1280 \times1024$ $pixels$ for movie 2. 

\section{Figures}
\begin{figure*} [h]
\renewcommand{\figurename}{\textbf{C1}}
\renewcommand{\thefigure}{}
\includegraphics[scale=.75]{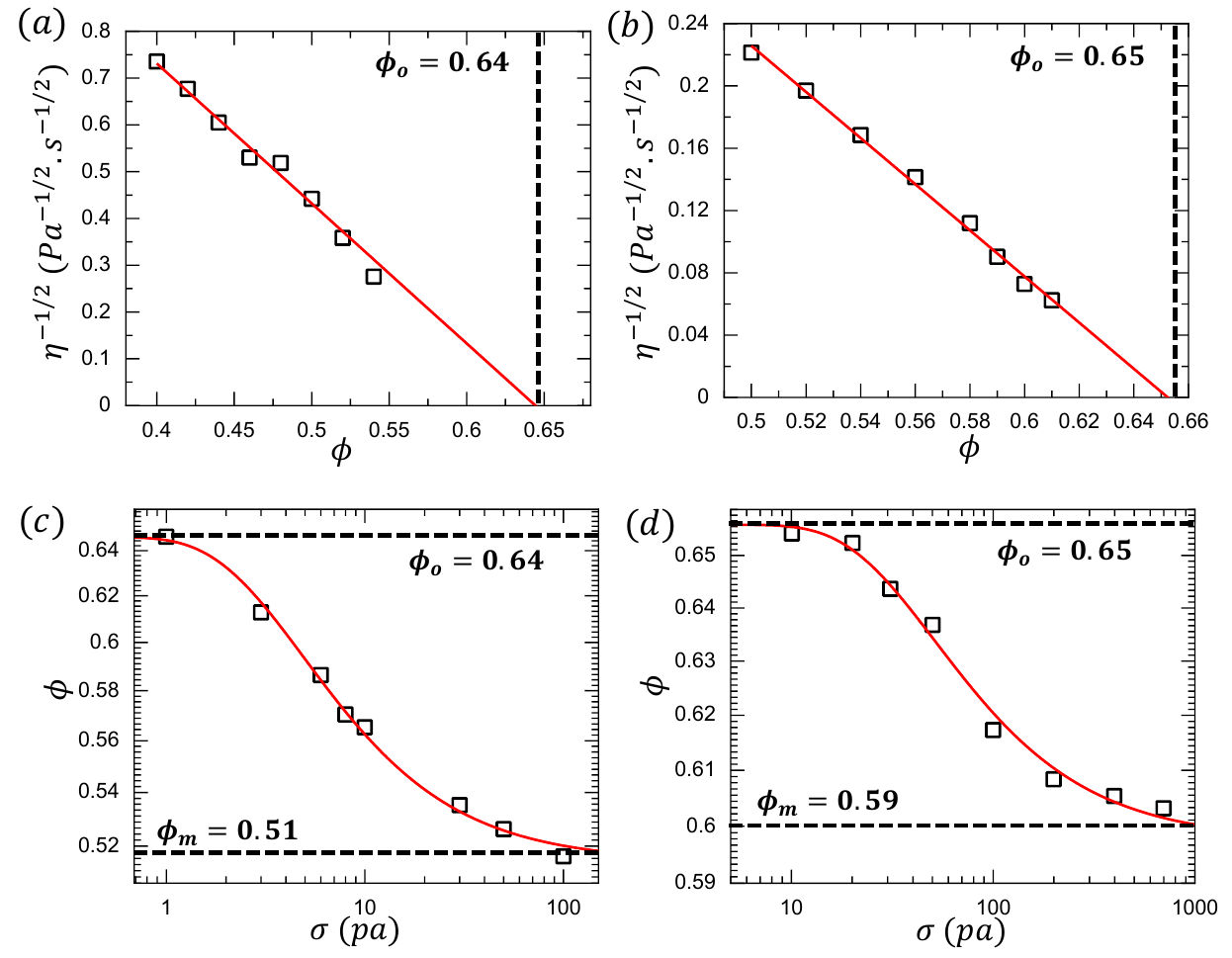}
\caption{\label{fig:wide} (a) and (b) Determination of $\phi_o$, as marked by vertical dashed lines, from the Krieger Dougherty (KD) relation fitting for cornstarch (CS) and polystyrene (PS) system respectively. (c) and (d) Variation of jamming volume fraction $\phi_J$ with $\sigma$ for CS and PS system. The red solid line indicates the fit to the Wyart-Cates(WC) model to get the parameters $\phi_m$ and $\sigma^*$. The top and bottom horizontal dashed lines indicate the obtained $\phi_0$ and $\phi_m$ values. For CS system $\phi_0$ =0.64 , $\phi_m$=0.51  and $\sigma^* =4.5$ $Pa$  ; for PS system $\phi_0  = 0.65$, $\phi_m$ = 0.59  and $\sigma^* =59$ $Pa$.}
\end{figure*}
\begin{figure*}
\renewcommand{\figurename}{\textbf{C2}}
\renewcommand{\thefigure}{}
\includegraphics[scale=0.58]{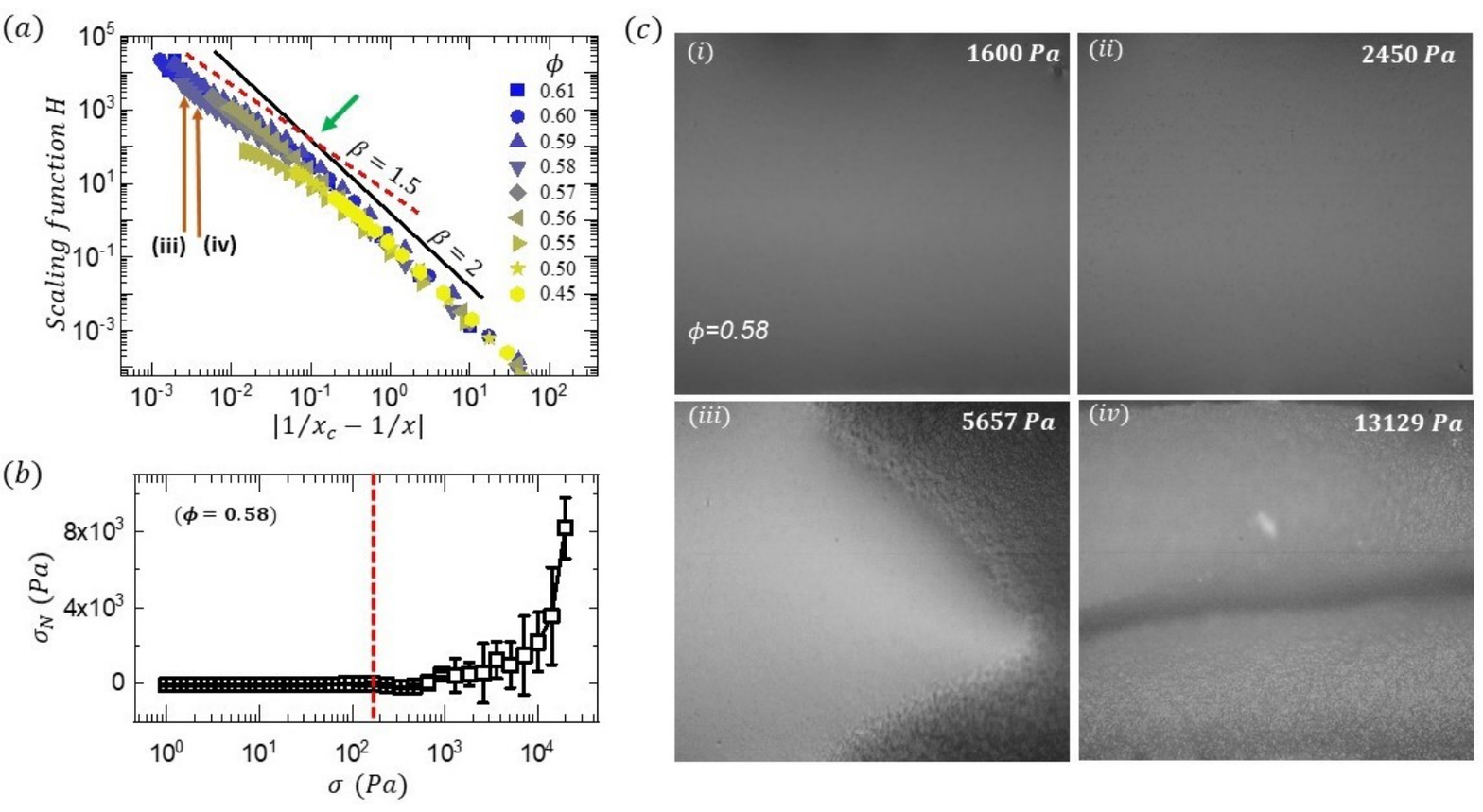}
     \caption{(a) Scaling curves for dense suspension of small polystyrene particles of diameter $d=2.65 \pm0.13$ $\mu m$ dispersed in PEG 400 at different volume fractions as shown in the legend. The black solid line and red dashed line represent the slope $-2$ and $-1.5$ respectively. The crossover point is marked by the green arrow. (b) Variation of normal stress $\sigma_N$ with shear stress $\sigma$ during flow curve for small polystyrene system ($\phi=0.58$). The corresponding cross-over stress $\sigma_c = 167 Pa$ is marked by the vertical dashed line. (c) Sample boundary images at four different stress values for $\phi=0.58 $. Panel (iii) and (iv) represents the onset of dilation (at $\sigma =5657$ $Pa$ ) and fracture (at $\sigma =13129$ $Pa$) respectively which are found deep inside the regime of slope $-1.5$ as shown in (a).}
\end{figure*}     
\begin{figure*}
\renewcommand{\figurename}{\textbf{C3}}
\renewcommand{\thefigure}{}
\centering
\includegraphics[scale=0.5]{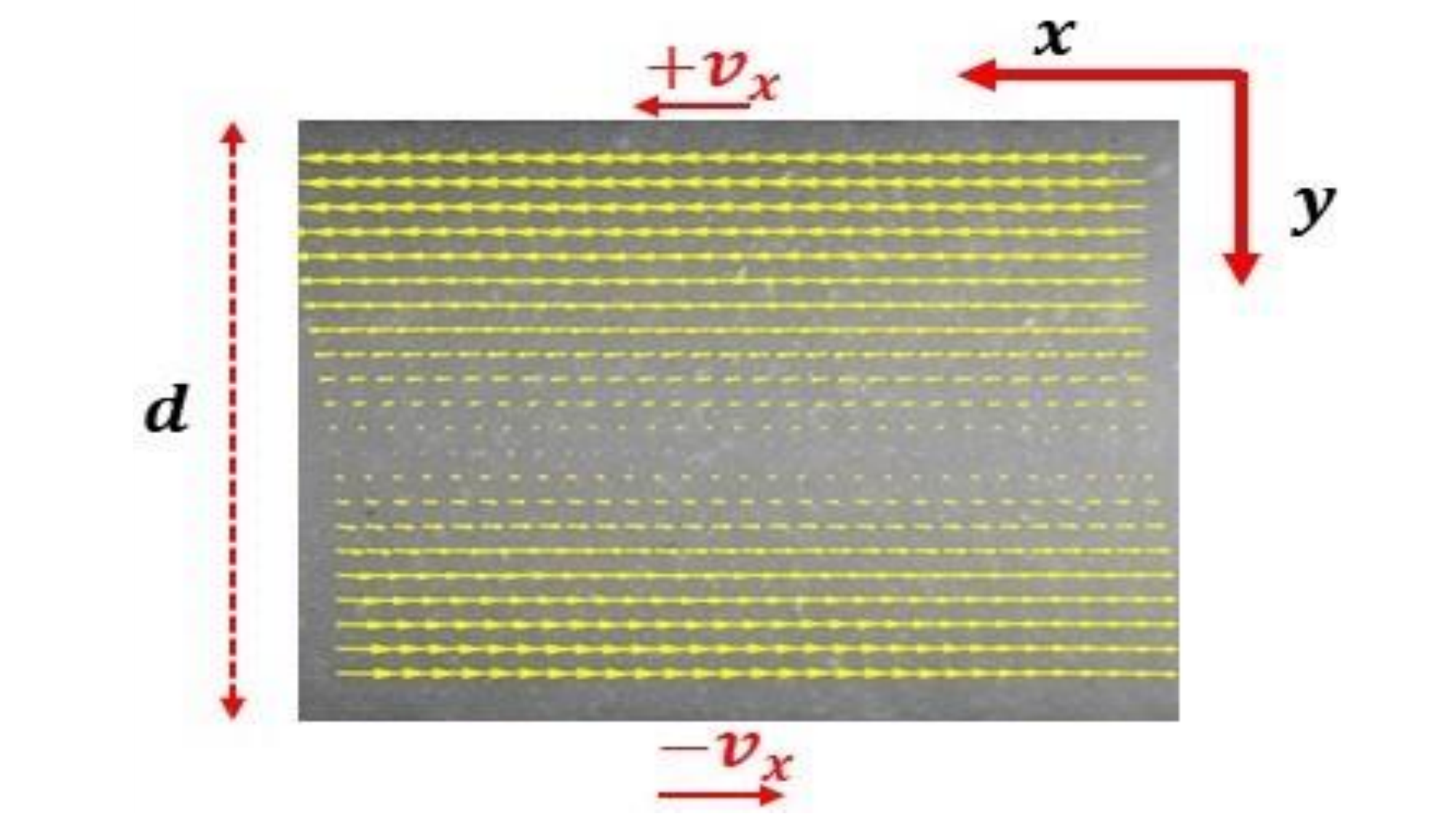}
\caption{\label{fig:wide}Typical PIV window for sample boundary image with top and bottom plate moving at velocity $+v_x$ and $-v_x$ respectively. Yellow arrows represent the velocity vector parallel to the plate (obtained from PIV).  ‘d’ is the gap between the two plates.}
\end{figure*}

\end{document}